\documentstyle [12pt] {article}
\input epsf
\newcommand{\gtwid}{\mathrel{\raise.3ex\hbox{$>$\kern-.75em\lower1ex
\hbox{$\sim$}}}}
\newcommand{\ltwid}{\mathrel{\raise.3ex\hbox{$<$\kern-.75em\lower1ex
\hbox{$\sim$}}}}
\newcommand{\beq}{\begin{equation}}
\newcommand{\eeq}{\end{equation}}
\newcommand{\beqs}{\begin{eqnarray}}
\newcommand{\eeqs}{\end{eqnarray}}

\catcode`@=11
\@addtoreset{equation}{section}
\@addtoreset{equation}{subsection}
\def\theequation{\ifnum\value{section}=0 \arabic{equation}\ignorespaces
\else \ifnum\value{section}=-1 A.\arabic{equation}\ignorespaces
\else \ifnum\value{subsection}=0 \thesection.\arabic{equation}\ignorespaces
\else \thesection.\arabic{subsection}.\arabic{equation}\ignorespaces
                           \fi
                      \fi
                 \fi}
\catcode`@=12

\begin{document}

\def\thefootnote{\fnsymbol{footnote}}
\baselineskip 6.0mm

\begin{flushright}
\begin{tabular}{l}
ITP-SB-96-42 
\end{tabular}
\end{flushright}

\vspace{4mm}
\begin{center}

{\bf Asymptotic Limits and Zeros of Chromatic Polynomials and }

\vspace{2mm}

{\bf Ground State Entropy of Potts Antiferromagnets} 

\vspace{8mm}

\setcounter{footnote}{0}
Robert Shrock\footnote{email: shrock@insti.physics.sunysb.edu}
\setcounter{footnote}{6}
and Shan-Ho Tsai\footnote{email: tsai@insti.physics.sunysb.edu}

\vspace{6mm}

Institute for Theoretical Physics  \\
State University of New York       \\
Stony Brook, N. Y. 11794-3840  \\

\vspace{10mm}

{\bf Abstract}
\end{center}
    We study the asymptotic limiting function 
$W(\{G\},q) = \lim_{n \to \infty}P(G,q)^{1/n}$, where $P(G,q)$ is the chromatic
polynomial for a graph $G$ with $n$ vertices.  We first discuss a subtlety in 
the definition of $W(\{G\},q)$ resulting from the fact that at certain special
points $q_s$, the following limits do not commute: 
$\lim_{n \to \infty} \lim_{q \to q_s} P(G,q)^{1/n} \ne
\lim_{q \to q_s} \lim_{n \to \infty} P(G,q)^{1/n}$. 
We then present exact calculations of $W(\{G\},q)$ and determine the 
corresponding analytic structure in the complex $q$ plane for a 
number of families of graphs $\{G\}$, including circuits, wheels, biwheels, 
bipyramids, and (cyclic and twisted) ladders.  We study the zeros of the
corresponding chromatic polynomials and prove a theorem that for certain
families of graphs, all but a finite number of the zeros lie exactly on a 
unit circle, whose position depends on the family.  Using the connection of 
$P(G,q)$ with the zero-temperature Potts antiferromagnet, we derive a 
theorem concerning the maximal finite real point of non-analyticity in
$W(\{G\},q)$, denoted $q_c$ and apply this theorem to deduce that 
$q_c(sq)=3$ and $q_c(hc) = (3+\sqrt{5})/2$ for the square and honeycomb 
lattices.  Finally, numerical calculations of $W(hc,q)$ and $W(sq,q)$ are 
presented and compared with series expansions and bounds. 

\vspace{16mm}

\pagestyle{empty}
\newpage

\pagestyle{plain}
\pagenumbering{arabic}
\renewcommand{\thefootnote}{\arabic{footnote}}
\setcounter{footnote}{0}

\section{Introduction}

   An important question in graph theory is the following: using $q$ different
colors, what is the number of ways $P(G,q)$ in which one can color a graph 
$G$, having $n$ vertices, such that no two adjacent vertices have the same 
color?  The function $P(G,q)$, first introduced by Birkhoff \cite{birk}, is
a polynomial in $q$ of order $n$ and has been the subject of mathematical study
for many years \cite{whit}-\cite{bl}; reviews are
Refs. \cite{rtrev},cite{graphs} . 
Clearly, a general upper bound on a chromatic polynomial is $P(G,q) \le q^n$,
since the right-hand side is the number of ways that one can color the
$n$-vertex graph $G$ without any constraint.  Consequently, it is of interest
to study the limiting function 
\beq
W(\{G\},q) = \lim_{n \to \infty} P(G,q)^{1/n}
\label{w}
\eeq
where the symbol $\{G\}$ denotes the limit as $n \to \infty$ of the family of
$n$-vertex graphs of type $G$. 

   This limit has some characteristics in common with the thermodynamic limit 
in statistical mechanics, in which one defines a partition function at a 
given temperature $T$ and external field $H$ as 
$Z = \sum_{\{ \sigma_i \} }e^{-\beta {\cal H}}$ (where the
Hamiltonian ${\cal H}$ describes the interactions of the spins $\sigma_i$, 
and $\beta =(k_BT)^{-1}$) and then, starting with a finite, usually regular, 
$d$-dimensional $n$-vertex lattice $G = \Lambda$ with some specified boundary 
conditions, one considers the reduced free energy (per site) $f$ in the 
thermodynamic limit, 
\beq
e^f = \lim_{n \to \infty} Z^{1/n}
\label{f}
\eeq
(Here, $f$ is related to the actual free energy $F$ by $f=-\beta F$.) 
For spin models in the physical temperature range, $0 \le \beta \le \infty$,
the partition function $Z$ is positive.  In the case of the chromatic
polynomial, for sufficiently large $q$, $P(G,q) > 0$.  In both cases, one  
naturally chooses the real positive $1/n$'th roots in the respective
eqs. (\ref{w}) and (\ref{f}).  

    Although the number of colors $q$ is an integer in the initial 
mathematical definition of the chromatic polynomial, one may generalize 
$q$ to a real or, indeed, complex variable.  We shall consider this
generalization here and study the function $W(\{G\},q)$ and the related zeros
of $P(G,q)$ in the complex $q$ plane for various families of graphs $G$.  
For certain ranges of real $q$, $P(G,q)$ can be negative, and, of course, 
when $q$ is complex, so is $P(G,q)$ in general. In these cases it may not be 
obvious, {\it a priori}, which of the $n$ roots 
\beq
P(G,q)^{1/n} = \{ |P(G,q)|^{1/n}e^{2\pi i r/n} \} \ , \quad r=0,1,...,n-1
\label{pphase}
\eeq
to choose in eq. (\ref{w}).  
Consider the function $W(\{G\},q)$ defined via eq. (\ref{w})
starting with $q$ on the positive real axis where $P(G,q) > 0$, and consider
the maximal region in the complex $q$ plane which can be reached by analytic 
continuation of this function.  We denote this region as $R_1$.  Clearly, the
phase choice in (\ref{pphase}) for $q \in R_1$ is that given by $r=0$, namely 
$P(G,q)^{1/n} = |P(G,q)|^{1/n}$. However, as we shall see via exactly solved
cases, there can also be families of graphs $\{G\}$ for which the analytic
structure of $W(\{G\},q)$ includes other regions not analytically connected to
$R_1$, and in these regions, there may not be any canonical choice of phase 
in (\ref{pphase}).  We shall discuss this further below. 

   Besides being of interest in mathematics, chromatic polynomials $P(G,q)$ 
and their asymptotic limits $W(\{G\},q)$ have a deep connection with 
statistical mechanics, specifically, the Potts antiferromagnet (AF)
\cite{potts}-\cite{wurev}.  Denote the partition function for the
(isotropic, nearest-neighbor, zero-field) $q$-state Potts model 
at a temperature $T$ as 
$Z = \sum_{ \{ \sigma_n \} } e^{-\beta {\cal H}}$ with the Hamiltonian
\beq
{\cal H} = -J \sum_{\langle ij \rangle} \delta_{\sigma_i \sigma_{j}}
\label{ham}
\eeq
where $\sigma_i=1,...,q$ are $Z_q$-valued variables on each site $i \in
\Lambda$. Define
\beq
K = \beta J \ , \qquad a = e^K
\label{a}
\eeq
For the Potts antiferromagnet ($J < 0$), in the limit 
$T \to 0$, i.e., $K \to -\infty$, 
the partition function only receives nonzero contributions from spin
configurations in which $\sigma_i \ne \sigma_j$ for nearest-neighbor vertices
$i$ and $j$, and hence, formally, 
\beq
Z(\Lambda,q,K=-\infty) = P(\Lambda,q)
\label{zprel}
\eeq
whence 
\beq
\exp \Bigl ( f(\Lambda,q,K=-\infty) \Bigr ) = W(\Lambda,q)
\label{fwrel}
\eeq
where $G=\Lambda$ denotes the lattice.  

  However, as we shall discuss in detail, the limit (\ref{w}), and hence the
resultant function $W(\{G\}=\Lambda,q)$ is not well-defined at certain special 
points $q_s$ without specifying further information.  This constitutes a
fundamental difference between the limits (\ref{w}) and (\ref{f}); in 
statistical mechanics, if the point $K_0$ lies within the interior of a given
physical phase, then $f(\Lambda,K)$ is a (real) analytic function of $K$. 
Furthermore, in statistical mechanics the limit $K \to K_0$ for a physical $K$,
and the thermodynamic limit $n \to \infty$ (with the $d$-dimensional volume
$vol_d(\Lambda) \to \infty$) commute \cite{bc}
\beq
\lim_{n \to \infty} \lim_{K \to K_0} Z^{1/n} = 
\lim_{K \to K_0} \lim_{n \to \infty} Z^{1/n}
\label{zcomm}
\eeq
These limits still commute for complex $K$
In contrast, the definition of $W(\Lambda,q)$ involves a further subtlety, 
since at certain special points $q_s$ the following limits do {\it not} 
commute (for any choice of $r$ in eq. (\ref{pphase})):
\beq
\lim_{n \to \infty} \lim_{q \to q_s} P(G,q)^{1/n} \ne 
\lim_{q \to q_s} \lim_{n \to \infty} P(G,q)^{1/n}
\label{wnoncomm}
\eeq
As we shall discuss, the origin of this noncommutivity of limits 
is an abrupt change in the behavior of $P(G,q)$ in the vicinity of such a 
point $q_s$; for $q \ne q_s$, $P(G,q)$ grows exponentially as the number of 
vertices $n$ in $G$ goes to infinity: $P(G,q) \sim a^n$ for some nonzero $a$, 
whereas precisely at $q=q_s$, it has a completely different type of behavior, 
which, in all of the cases considered here is $P(G,q_s) = c_0(q_s)$ where 
$c_0(q)$ may either be a constant, independent of $n$ or may depend on $n$ in a
way that does not involve exponential growth, such as $(-1)^n$. 
The set of special points $\{q_s\}$ 
includes $q=0$, $q=1$, and, on any graph $G$ which contains at least one 
triangle, also $q=2$; at these points, $P(G,q_s)=0$.  It is also possible for 
$P(G,q_s)$ to be equal to a nonzero constant at $q_s$.  We shall discuss this
further in Section 2. 

   Before proceeding, we mention that, in addition to
Refs. \cite{birk}-\cite{rtrev}, some relevant previous works are Refs. 
\cite{baxter70}-\cite{read91}. 

    This paper is organized as follows.  In Section 2 we discuss a subtlety in
the definition of $W(\{G\},q)$, and in Section 3 we present some general
results on the analytic structure of this function in the complex $q$ plane. In
Section 4 we give a number of exact solutions for $W(\{G\},q)$ for various 
families of graphs $\{G\}$ and calculate the resultant diagrams, showing
the analytic structure of $W(\{G\},q)$ in the complex $q$ plane.  
Section 5 contains a discussion of zeros of chromatic polynomials for various
families of graphs and, in particular, a theorem on the location
and density of zeros of $P(G,q)$ for certain families $G$. In Section 6 we 
present a theorem specifying the maximal value of $q$, for a given lattice
$\Lambda$,  where the region boundary ${\cal B}$ for $W(\Lambda,q)$ crosses 
the real $q$ axis, and we apply this to specific lattices.  Section 7 gives
numerical calculations of $W(\Lambda,q)$ for the honeycomb and square
lattices and a comparison with large-$q$ series. Section 8 contains some 
concluding remarks. 

\section{Definition of $W(\{G\},q)$ }

   In order to discuss the subtlety in the definition of $W(\{G\},q)$, we first
recall the following general properties concerning the zeros of chromatic 
polynomials.  First, for any graph $G$, 
\beq
P(G,q=0) = 0 
\label{pq0}
\eeq
Second, for any graph $G$ consisting of at least two vertices (with a bond
connecting them), 
\beq
P(G,q=1) = 0
\label{pq1}
\eeq
Third, for any graph $G$ containing at least one triangle, 
\beq
P(G,q=2) = 0 \quad {\rm if} \quad G \supseteq \triangle 
\label{pq2}
\eeq
These properties are obvious from the definition of $P(G,q)$, given that 
one must color adjacent vertices with different colors.  Since $P(G,q)$ is a
polynomial, each of these zeros at the respective values $q_0=0,1$ or 2 means
that $P(G,q)$ must factorize according to 
\beq
P(G,q) = (q-q_0)^{b(q_0)}Q(G,q)
\label{pfac}
\eeq
where $b(q_0)$ is a positive integer and $Q(G,q_0) \ne 0$.  One may distinguish
two particular cases that occur for cases we have studied: 
(i) $b(q_0) = b_0 + b_1n$; (ii) $b(q_0) = b_0$, where 
$b_0$ and $b_1$ are integers independent of $n$.  In the first, case, 
\beq
W(\{G\},q) = (q-q_0)^{b_1}\lim_{n \to \infty}Q(\{G\},q)^{1/n}
\label{wqscale}
\eeq
so that the two different orders of limits in (\ref{wnoncomm}) do commute.
This type of behavior is observed for tree graphs, our first example below.
However, in all of the other cases which we have studied, the second type of
behavior (ii) holds.  Hence for these families of graphs, as a consequence of
the basic fact that 
\beq
\lim_{n \to \infty} x^{1/n} = \cases{1 & if $x \ne 0$ \cr
                                     0 & if $x=0$ \cr}
\label{basiclim}
\eeq
and hence, 
$\lim_{n \to \infty} \lim_{q \to q_0} (q-q_0)^{b(q_0)/n} = 0$ and 
$\lim_{q \to q_0} \lim_{n \to \infty} (q-q_0)^{b(q_0)/n} = 1$, 
the noncommutativity of the limits in eq. (\ref{wnoncomm}) follows (for any
value of $r$ in (\ref{pphase})): 
\beq
\lim_{n \to \infty} \lim_{q \to q_0} P(G,q)^{1/n} = 0
\label{pq0nq}
\eeq
whereas 
\beq
\lim_{q \to q_0} \lim_{n \to \infty} P(G,q)^{1/n} = \lim_{q \to q_0} \lim_{n
\to \infty} Q(G,q)^{1/n} \ne 0
\label{pq0qn}
\eeq

  More generally, eq. (\ref{w}) is insufficient to define $W(\{G\},q)$ not
just in the vicinity of a zero of $P(G,q)$, but also in the vicinity of any
special point $q_s$ where the asymptotic behavior of $P(G,q)$ changes abruptly
 from 
\beq
P(G,q) \sim a^n \quad {\rm as} \quad  n \to \infty
\label{pgasym}
\eeq
with $a$ a nonzero constant, to 
\beq
P(G,q_s) = const. \quad {\rm as} \quad n \to \infty
\label{pgqs}
\eeq
In case (ii) of eq. (\ref{pfac}), with $b(q_s)= b_0$, one encounters this type
of abrupt change in behavior with the constant in eq. (\ref{pgqs}) equal to
zero.  This is the origin of the noncommutativity of limits 
in eq. (\ref{wnoncomm}) at
$q=0,1$ and, for $G \supseteq \triangle$, at $q=2$.  However, this
noncommutativity is more general and can also occur when the constant in
eq. (\ref{pgqs}) is nonzero. An example is provided by the point $q_s=3$ on 
the triangular lattice; there are $3! = 6$ ways of coloring a triangular
lattice graph (with the technical provision that for finite triangular lattice
graphs, one uses boundary conditions which do not introduce frustration).
Denoting such a triangular lattice graph as $tri_n$, it follows that 
$P(tri_n,3)=6$, which is of the form of eq. (\ref{pgqs}) with a nonzero
constant.  For such cases, where the constant in eq. (\ref{pgqs}) is nonzero,
one has 
\beq
\lim_{n \to \infty} \lim_{q \to q_s} |P(G,q)|^{1/n} = 1
\label{pnqs1}
\eeq
while 
\beq
\lim_{q \to q_s} \lim_{n \to \infty} |P(G,q)|^{1/n} = |a|
\label{pqsna}
\eeq
where, in general, $a \ne 1$.  Finally, the set of points $\{q_s\}$
also may include a continuous set comprising part of a region boundary, as will
be discussed in theorem 1, part (e) below.

   Because of the noncommutativity (\ref{wnoncomm}), 
the formal definition (\ref{w}) is, in 
general, insufficient to define $W(\{G\},q)$ at the set of special points
$\{q_s\}$; at these points, one must also specify the order of the limits in 
(\ref{wnoncomm}).  One can maintain the analyticity of $W(\{G\},q)$ at these
special points $q_s$ of $P(G,q)$ by choosing the order of limits in the 
right-hand side of eq. (\ref{wnoncomm}): 
\beq
W(\{G\},q_s)_{D_{qn}} \equiv \lim_{q \to q_s} \lim_{n \to \infty} P(G,q)^{1/n} 
\label{wdefqn}
\eeq
As indicated, we shall denote this definition as $D_{qn}$, where the 
subscript indicates the order of the limits.  Although this definition 
maintains the analyticity of $W(\{G\},q)$ at the special points $q_s$, it 
produces a function $W(\{G\},q)$ whose values at the points $q_s$ differ 
significantly from the values which one would get for $P(G,q_s)^{1/n}$ with 
finite-$n$ graphs $G$.  The definition based on the opposite order of limits, 
\beq
W(\{G\},q_s)_{D_{nq}} \equiv \lim_{n \to \infty} \lim_{q \to q_s} P(G,q)^{1/n}
\label{wdefnq}
\eeq
gives the expected results like $W(G,q_s)=0$ for $q_s=0,1$, 
and, for $G \supseteq \triangle$, $q=2$, as well as $W(tri_n,q=3)=1$, but
yields a function $W(\{G\},q)$ with discontinuities at the set of points
$\{q_s\}$.  In our results below, in order to avoid having to write special
formulas for the points $q_s$, we shall adopt the definition $D_{qn}$ but at
appropriate places will take note of the noncommutativity of limits 
(\ref{wnoncomm}).

   As noted in the introduction, the noncommutativity of limits
(\ref{wnoncomm}) and resultant subtlety
in the definition of $W(\{G\},q)$ is fundamentally different from the behavior
of the (otherwise somewhat analogous) function $e^{f(G,K)}$ in statistical 
mechanics (where for this discussion, we consider a general statistical
mechanical model and its reduced free energy, $f$, and do not restrict to the
Potts model).  The set $\{q_s\}$ includes certain discrete points lying within
regions in the complex $q$ plane 
where $W(\{G\},q)$ is otherwise an analytic function. 
Now, considering the thermodynamic limit of a statistical mechanical
model on a lattice $G = \Lambda$, one knows that for physical $K$, after the 
additive term $(\zeta/2)K + h$ is removed (where $\zeta$ is the coordination
number of the lattice $\Lambda$, $h = \beta H$, and this removes
the trivial isolated infinities in $f$ at $K=\infty$ and $h=\infty$), i.e. 
after defining $f(\Lambda,K) = (\zeta/2)K+h+f_r(\Lambda,K)$, the function 
$f_r(\Lambda,K)$ is 
analytic within the interior of a given phase.
We should remark that in our studies of
the properties of spin models generalized to complex-temperature, we have 
established that there may be singularities in thermodynamic quantities in the
interiors of (complex-temperature extensions of physical) phases; specifically,
we proved a theorem (theorem 6 in Ref. \cite{cmo}) that on a lattice with odd
coordination number, the zero-field 
Ising model partition function vanishes, and the 
free energy $f$ has a negatively divergent singularity, at the
complex-temperature point $z=-1$, where $z=e^{-2K_I}$ and $K_I=\beta J_I$ 
is the 
Ising spin-spin coupling.  For the honeycomb lattice, $z=-1$ lies on a phase
boundary \cite{chitri}, but for the heteropolygonal lattice denoted 
$3 \cdot 12^2$, $z=-1$ lies in the interior of the complex-temperature
extension of the ferromagnetic phase \cite{cmo}.  However, in the quantity
analogous to $W(\Lambda,q)$, namely, $e^{f(\Lambda,K)}$, this singularity is a 
zero, not a discontinuity and furthermore it is not associated with the type of
noncommutativity analogous to (\ref{wnoncomm}).  
The reason for this is that when one factorizes the
(zero-field) partition function in a manner similar to eq. (\ref{pfac}), 
\beq
Z(\Lambda,z) = (z+1)^{n/2}Z_s(\Lambda,z)
\label{zz}
\eeq
(which defines $Z_s(\Lambda,z)$), 
the exponent of the factor $(z+1)$ is proportional to $n$, as in case (i) for
eq. (\ref{pfac}), so that the following limits commute:
\beq
\lim_{z \to -1} \lim_{n \to \infty} Z(\Lambda_n,z) = 
\lim_{n \to \infty} \lim_{z \to -1} Z(\Lambda_n,z) = 0
\label{zcommz}
\eeq

One should also contrast the noncommutativity (\ref{wnoncomm}) with the very
different type of noncommutativity which applies to a symmetry-breaking order
parameter such as a (uniform or staggered) magnetization in a statistical
mechanical spin model (above its lower critical dimensionality, so that it has
a symmetry-breaking phase transition).  
Here, in both the symmetric, high-temperature phase and
the low-temperature phase with spontaneously broken symmetry, if one removes
the external field before taking the thermodynamic limit, the magnetization 
vanishes:
\beq
\lim_{n \to \infty} \lim_{H \to 0} M(\Lambda,K,H) = 0
\label{mnh}
\eeq
whereas in the low-temperature, symmetry-broken phase ($K > K_c)$, there is a 
nonzero magnetization in the thermodynamic limit: 
\beq
\lim_{H \to 0} \lim_{n \to \infty} M(\Lambda,K,H) \ne 0 \ , \quad {\rm for}
\quad K > K_c
\label{mhn}
\eeq
However, this noncommutativity is quite different from that in
eq. (\ref{wnoncomm}):  this is clear from the fact that, among other things, 
(\ref{wnoncomm}) can occur at a discrete, isolated set of special points 
$q_s$ (as well as possibly a continuous set on a region boundary of type (e) in
theorem 1 below), whereas the
noncommutativity in eqs. (\ref{mnh}), (\ref{mhn}) occurs throughout the
low-temperature, broken-symmetry phase of the spin model and, indeed, can be
used to characterize this phase, with the spontaneous magnetization $M(K,0)$
(defined of course by the second ordering of limits, (\ref{mhn})) constituting
the order parameter.

\section{Analytic Structure of $W(\{G\},q)$ }

   As noted, we shall consider the variable $q$ to be extended from the
positive integers to the complex numbers.  Although for a given graph $G$, 
$P(G,q)$ is a polynomial and hence, {\it a fortiori}, is an analytic function 
of $q$, the function $W(\{G\},q)$ which describes the $n \to \infty$ limit of a
given family of graphs $\{G\}$ will, in general, fail to be analytic at certain
points.  These points may form a discrete or continuous set; if the set is
continuous, it may separate certain regions of the complex $q$ plane, which we
denote $R_i$.  We shall denote the boundary separating regions $R_i$ and 
$R_j$ as ${\cal B}(R_i,R_j)$ and the union of all components of regional 
boundaries as ${\cal B} = \bigcup_{i,j} {\cal B}(R_i,R_j)$.  On these
boundaries, $W(\{G\},q)$ is non-analytic.  We shall illustrate this with exact
results below.  These regions are somewhat similar to complex-temperature
extensions or complex-field extensions of physical phases in statistical 
mechanical models.  However, there are also some fundamental differences.  One
of these is the noncommutativity of limits 
discussed in the previous section.  Another is
that in the case of $W(\{G\},q)$, it is not clear what would play the role of
the physical concept of a order parameter characterizing a given phase 
(and its complex extension, such as to complex-temperature).  Therefore, we
shall use the terms ``region'' and ``region diagram'' rather than (complex
extensions of) ``phase'' and ``phase diagram''. 

    A question that arises when one considers the somewhat related regions
(phases) in complex-temperature or complex-field variables for statistical 
mechanical spin models concerns the dimensionality of the locus of points 
where the reduced free energy $f$ is non-analytic.  Where the premise of the
Yang-Lee theorem \cite{yl,ly} holds (i.e., for physical temperature and 
Hamiltonians with ferromagnetic, 
but not necessarily nearest-neighbor, two-spin interactions, 
${\cal H} = -\sum_{\langle i j \rangle} \sigma_i J_{ij} \sigma_j - 
H\sum_i \sigma_i$ on arbitrary graphs), it states that the zeros of $Z$ 
in the complex $\mu = e^{-2\beta H}$ plane lies on the unit circle $|\mu|=1$
and hence, in the thermodynamic limit where these merge to form the continuous
locus of points where $f$ is non-analytic in the $\mu$ plane, this locus is 
one-dimensional.  In the case of complex-temperature, taking the
Ising model for illustration, the locus of points in the $z=e^{-2K}$ plane is 
usually one-dimensional for isotropic spin-spin couplings $J$, but on the
heteropolygonal $4 \cdot 8^2$ lattice, it fills a two-dimensional area in this
plane even for isotropic couplings \cite{cmo}.  In all of the exact results for
$W(\{G\},q)$ that we shall present below, the dimension of the continuous locus
of points where $W(\{G\},q)$ is non-analytic, is 
$\dim \{ {\cal B} \} = 1$.  

   We next present a general theorem. 

\begin{flushleft}

Theorem 1 

\end{flushleft}

  Let $G$ be a graph with $n$ vertices and suppose that $P(G,q)$ has the 
form 
\beq
P(G,q) = q(q-1)\Bigl \{  c_0(q) + \sum_{j=1}^{N_a} c_j(q)a_j(q)^n \Bigr \}
\label{pgsum}
\eeq
where $c_j(q)$ are polynomials in $q$.  Here $c_0(q)$ may contain $n$-dependent
terms, such as $(-1)^n$, but does not grow with $n$ like $a^n$. 
This form and the additional 
factorization (\ref{pgsumtri}) are motivated by the exact solutions to be 
presented below. Note that the fact that $P(G,q)$ is a polynomial 
guarantees that, for a given $G$, $N_a$ is finite. For a fixed $q$ and, more 
generally,
for a given region in the complex $q$ plane, we define a term $a_\ell(q)$ to be
leading if for $q$ in this region $|a_\ell(q)| \ge 1$ and 
$|a_\ell(q)| > |a_j(q)|$ for all $j \ne \ell$.  
Without loss of generality, we can write (\ref{pgsum}) so that
the $a_j(q)$ are different functions of $q$. 
Then our theorem states that (a) if $N_a \ge 2$ and there exists some $\ell$
such that $|a_\ell(q)| > 1$ in a given region of the complex $q$ plane, then 
if in this region, $|a_j(q)| < 1$, the term $a_j(q)$ does not contribute to 
the limiting function $W(G,q)$; (b) if $N_a=1$ and $c_0(q) \ne 0$, then if 
$|a_1| < 1$, this term again does not contribute to $W(G,q)$, which is then 
determined by $c_0(q)$; (c) if $N_a \ge 1$ and $a_\ell(q)$ is a leading term 
in a given region of the $q$ plane then
(i) if this region is analytically connected to the positive real axis where 
$P(G,q) > 0$ so that $r=0$ in (\ref{pphase}), 
\beq
W(\{G\},q) = a_\ell(q) 
\label{wgq}
\eeq
while (ii) if (i) is not the case, then at least in terms of magnitudes, one
has the result 
\beq
|W(\{G\},q)| = |a_\ell(q)|
\label{wgqmag}
\eeq
(d) the regional boundaries ${\cal B}$ separating regions where different
leading terms dominate are determined by the degeneracy in magnitude of these 
leading terms: $|a_\ell(q)|=|a_{\ell'}(q)|$; (e) a regional boundary can also 
occur where one crosses from a region where there is a leading term
$a_\ell(q)$ to one where there is no leading term but there is a nonzero
$c_0(q)$ ; this type of boundary is given by the equation $|a_\ell(q)|=1$. 

\vspace{4mm}

\noindent Proof:  (a) is clear since if $|a_j(q)| < 1$, then $\lim_{n \to
\infty} a_j(q)^n = 0$, so in this limit it does not contribute to $W(G,q)$,
which is determined by the leading term, as specified in part (c).  Part (b)
follows by the same type of logic.  Note that if there is no $c_0(q)$ term and
if $N_a=1$ with $|a_1(q)| < 1$, then in this case, $a_1$ still determines
$W(G,q)$; an example of this is provided by the region $|q-1| < 1$ for the 
tree graphs $T_n$ to be discussed below.  Part (c) 
expresses the fact that in the limit $n \to \infty$, the contributions of
subleading terms are negligible relative to that of the leading term, and hence
the limiting function $W(\{G\},q)$ depends only on this leading term. As one 
moves from a region with one dominant term $a_\ell(q)$ to a region in which a
different term $a_{\ell'}(q) $ dominates, there is a non-analyticity 
in $W(\{G\},q)$ as it switches from $W(\{G\},q)=a_\ell(q)$ to 
$W(\{G\},q)=a_{\ell'}(q)$ for $r=0$ and similarly for nonzero $r$. 
This also proves (d).  Statement (e) follows in a similar way. 

    It is possible that $P(G,q)$ contains no term of the form $c_j(q)a_j(q)^n$
but instead only the term $c_0(q)$.  Moreover, in the case where $P(G,q)$
does contain such $c_j(q)a_j(q)^n$ terms, there may exist a region in the $q$
plane where $|a_j(q)| < 1$ for all $j=1,...,N_a$. In both of these cases, 
$W(\{G\},q)$ is determined by the remaining function $c_0(q)$. 

   If $G$ contains one or more triangles $\triangle$, then
 one may express $P(G,q)$ in the
form 
\beq
P(G,q)=q(q-1)(q-2)\Bigl \{ c_0(q)+\sum_{j=1}^{N_a}c_j(q)a_j(q)^n ) \Bigr \} 
\quad {\rm if} \quad G \supseteq \triangle
\label{pgsumtri}
\eeq
In this case, the same theorem applies, but with the further factorization
(\ref{pgsumtri}) taken into account.

   The subtlety in the definition of $W(\{G\},q)$ resulting from the
noncommutativity (\ref{wnoncomm}) is evident in the forms (\ref{pgsum}) and
(\ref{pgsumtri}).  With the definition (\ref{wdefqn}), if there is a leading
term $a_\ell$ at $q_0=0$, 1, and, for $\{G\} \supseteq \triangle$, also $=2$, 
then  
\beq
|W(\{G\},q_0)| = |a_\ell(q_0)|
\label{wq0}
\eeq
rather than zero, even though $P(G,q_0)=0$ at these points.  If there is no 
leading term in the vicinity of a given $q_0$, i.e., if $|a_j(q_0)|< 1$ for 
all $j=1,...,N_a$, then, if there is a $c_0(q)$ term, 
\beq
|W(\{G\},q)| = \lim_{q \to q_0} \lim_{n \to \infty} |c_0(q)|^{1/n} = 1
\label{wc0}
\eeq

\section{Exact Solutions for $W(\{G\},q)$} 

   In this section we calculate and discuss exact solutions for 
$W(\{G\},q)$ for various families $\{G\}$ of graphs.  We believe that these
give some interesting insights into the analytic properties of such 
functions and also into the exact results obtained by Baxter for the triangular
lattice.  Unless otherwise cited, chromatic polynomials can be found, 
together with further properties of graphs, in 
Refs. \cite{rtrev}-\cite{graphs}. 

\subsection{Tree Graphs}

   A tree graph $T_n$ is an $n$-vertex graph with no circuits and has the 
chromatic polynomial $P(T_n,q)=q(q-1)^{n-1}$.  Using the procedure discussed in
section 2, we choose $r=0$ in (\ref{pphase}) and obtain 
\beq
W(\{T\},q) = q-1 
\label{wt}
\eeq
This applies for all $q$; i.e., $W(\{T\},q)$ is analytic throughout the entire
(finite) complex $q$ plane. 

\subsection{Complete Graphs}

   An $n$-vertex graph is termed ``complete'' and denoted $K_n$ if each 
vertex is completely connected by bonds (edges) with all the other vertices. 
Thus, $K_3$ is the triangle, $K_4$ the tetrahedron, and so forth.  The
chromatic polynomial is $P(K_n,q) = \prod_{i=0}^{n-1} (q-i)$.  For a given $n$,
we may choose $r=0$ in (\ref{pphase}) by starting on the positive real $q$ axis
at a value $q > n-1$.  This yields 
\beq
W(\{K\},q) = 1 
\label{wk}
\eeq
With our definition (\ref{wdefqn}), $W(\{K\},q)$ is analytic in the entire 
(finite) $q$ plane. 
The zeros of $P(K_n,q)$ are comprised by the set $\{q_0\} = \{0,1,...,n-1\}$
and the noncommutativity of limits 
(\ref{wnoncomm}) occurs at each of these points. 

\subsection{Cyclic Graphs} 

\begin{figure}
\epsfxsize=3.5in
\epsffile{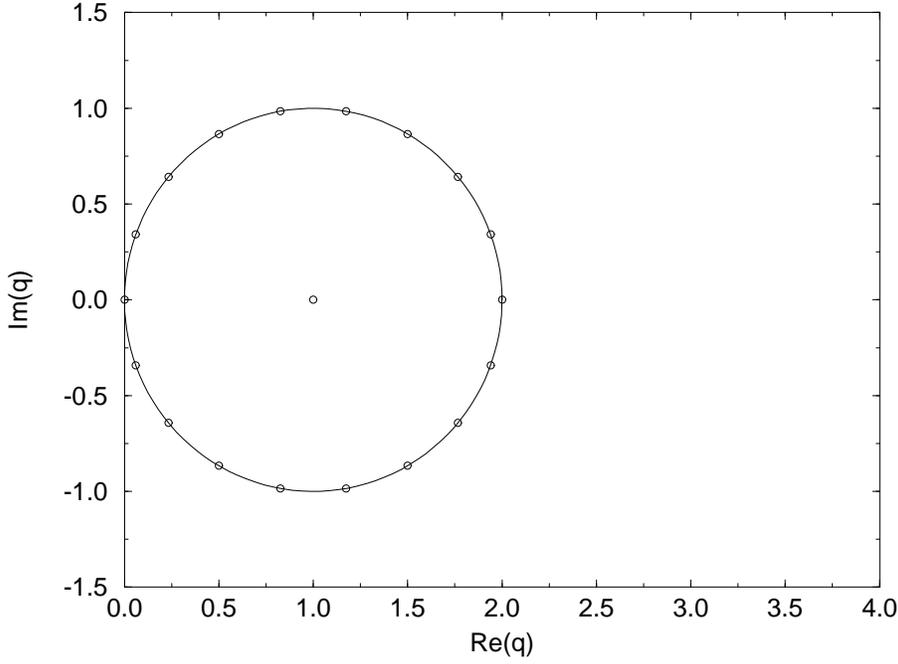}
\caption{Diagram showing regional boundaries comprising ${\cal B}$ for 
$W(\{C\},q)$ for cyclic graphs and zeros of $P(C_n,q)$ for $n=19$.} 
\label{cyclicfig}
\end{figure}

   For the cyclic graph $C_n$, i.e., the $n$-circuit, the chromatic polynomial
is \linebreak
$P(C_n,q) = (q-1)\Bigl \{ (q-1)^{n-1} + (-1)^n \Bigr \}$.  We find that the
analytic structure of $W(\{C\},q)$ differs in the two regions $R_1$ and $R_2$ 
consisting of $q$ satisfying $|q-1| > 1$ and $|q-1| < 1$, respectively.  The
boundary ${\cal B}$ separating these regions is thus the unit circle centered
at $q=1$.  These regions are shown in Fig. \ref{cyclicfig}.  We calculate 
\beq
W(\{C\},q) = q-1 \quad {\rm for} \quad q \in R_1
\label{wcr1}
\eeq
For $q \in R_2$, the first term in curly brackets, $(q-1)^{n-1} \to 0$ as 
$n \to
\infty$, and hence $P(C_n,q) \to (q-1)(-1)^n$.  This function does not have a
limit as $n \to \infty$.  However, we can observe that 
\beq
|W(\{C\},q)| = 1 \quad {\rm for} \quad q \in R_2
\label{wcr2}
\eeq
$|W(\{C\},q)|$ is, in general, discontinuous along ${\cal B}$.  For the choice
$r=0$ in (\ref{pphase}), it is continuous at $q=2$ and has a discontinuity of 
\beq
\lim_{q \searrow 0} W(\{C\},q) - \lim_{q \nearrow 0} W(\{C\},q) = 2
\label{cdisc}
\eeq
at $q=0$.  
In Fig. \ref{cyclicfig} we have also plotted the zeros of $P(C_n,q)$ for a 
typical value, $n=19$. These will be discussed in Section 5. 

\subsection{Wheel Graphs}

   The wheel graph $(Wh)_n$ is defined as an $(n-1)$-circuit $C_{n-1}$ with an
additional vertex joined to all of the $n-1$ vertices of $C_{n-1}$ (which can
be thought of as the center of the wheel), and is naturally defined for 
$n \ge 3$. We find that the diagram describing the analytic structure of
$W(\{Wh\},q)$ consists of the two regions $R_1$ and $R_2$ defined respectively
by $|q-2| > 1$ and $|q-2| < 1$, with the boundary ${\cal B}$ consisting of the
unit circle $|q-2|=1$.  (These $R_1$ and $R_2$ should not be
confused with the regions discussed in the previous subsection; we define the
regions $R_j$ differently for each family of graphs.) 
This diagram is thus similar to Fig. \ref{cyclicfig}
but with the circle moved one unit to the right; for brevity we do not
show it.  Using 
\beq
P((Wh)_n,q) = q(q-2)\{ (q-2)^{n-2} - (-1)^n \}
\label{pwn}
\eeq
we calculate 
\beq
W(\{Wh\},q) = q-2 \quad {\rm for} \quad q \in R_1 
\label{wwhn}
\eeq
For $q \in R_2$, since the first term in curly brackets in eq. (\ref{pwn}), 
$(q-2)^{n-2}$, goes to zero as $n \to \infty$, one is left only with the second
term, $-(-1)^n$; this term does not have a smooth limit as $n \to \infty$ and 
hence neither does the $1/n$'th power of this quantity.  However,
$|W(\{Wh\},q)|=1$ for $q \in R_2$.  Formally, one may choose the $1/n$'th 
root such that $W(\{Wh\},q) = -1$ for $q \in R_2$. 
The noncommutativity of limits in eq. (\ref{wnoncomm}) 
occurs at the discrete points $q=0,1,2$ and, for even $n$, also at $q=3$ since 
$P((Wh)_n, \ n \ {\rm even}, \ q=3)=0$.  More generally, for $q \ne 2$, the 
$n \to \infty$ limit is not well defined on the circle $|q-2|=1$. 

   One can also study wheel graphs with some spokes removed, which have been
of recent interest \cite{dl}.  Let us define the ``cut'' wheel 
$(cWh)_{n,\ell}$ as the $n$-vertex wheel graph with $\ell$ consecutive spokes
removed.  For example, from an analysis of the specific case $\ell=2$, we find
the same boundary ${\cal B}$, $|q-2|=1$, as for the asymptotic limit of the
wheel graphs, and, furthermore, $W(\{cWh\},q) = W(\{Wh\},q)$. 

\subsection{Biwheel Graphs}

  The biwheel graph $U_n$ is defined by adjoining a second vertex to all of the
other vertices in the wheel graph $(Wh)_{n-1}$ and is naturally defined for 
$n \ge 4$.  Here, we find that the diagram describing the analytic structure of
$W(\{U\},q)$ consists of the two regions $R_1$ and $R_2$ defined respectively
by $|q-3| > 1$ and $|q-3| < 1$ with ${\cal B}$ consisting of the circle
$|q-3|=1$.  The chromatic polynomial for these graphs is 
\beq
P(U_n,q) = q(q-1)(q-3)\Bigl \{ (q-3)^{n-3} +(-1)^n \Bigr \}
\label{pun}
\eeq
and from this we calculate 
\beq
W(\{U\},q) = q-3 \quad {\rm for} \quad q \in R_1 
\label{wun}
\eeq
For $q \in R_2$, the term 
$(q-3)^{n-3} \to 0$ as $n \to \infty$, and one is again left
with a discontinuous term $(-1)^n$.  As before, one has in general, that for $q
\in R_2$, $|W(\{U\},q)|=1$, and one can formally choose the 
$1/n$'th root such that $W(\{U\},q)=-1$ in this region. 
The noncommutativity of limits in (\ref{wnoncomm}) occurs at the discrete
points $q=0,1,2,3$, and also, if $n$ is odd, at $q=4$ since 
$P(U_n, \ n \ {\rm odd}, \ q=4)=0$.

\subsection{Bipyramid Graphs}

\begin{figure}
\epsfxsize=3.5in
\epsffile{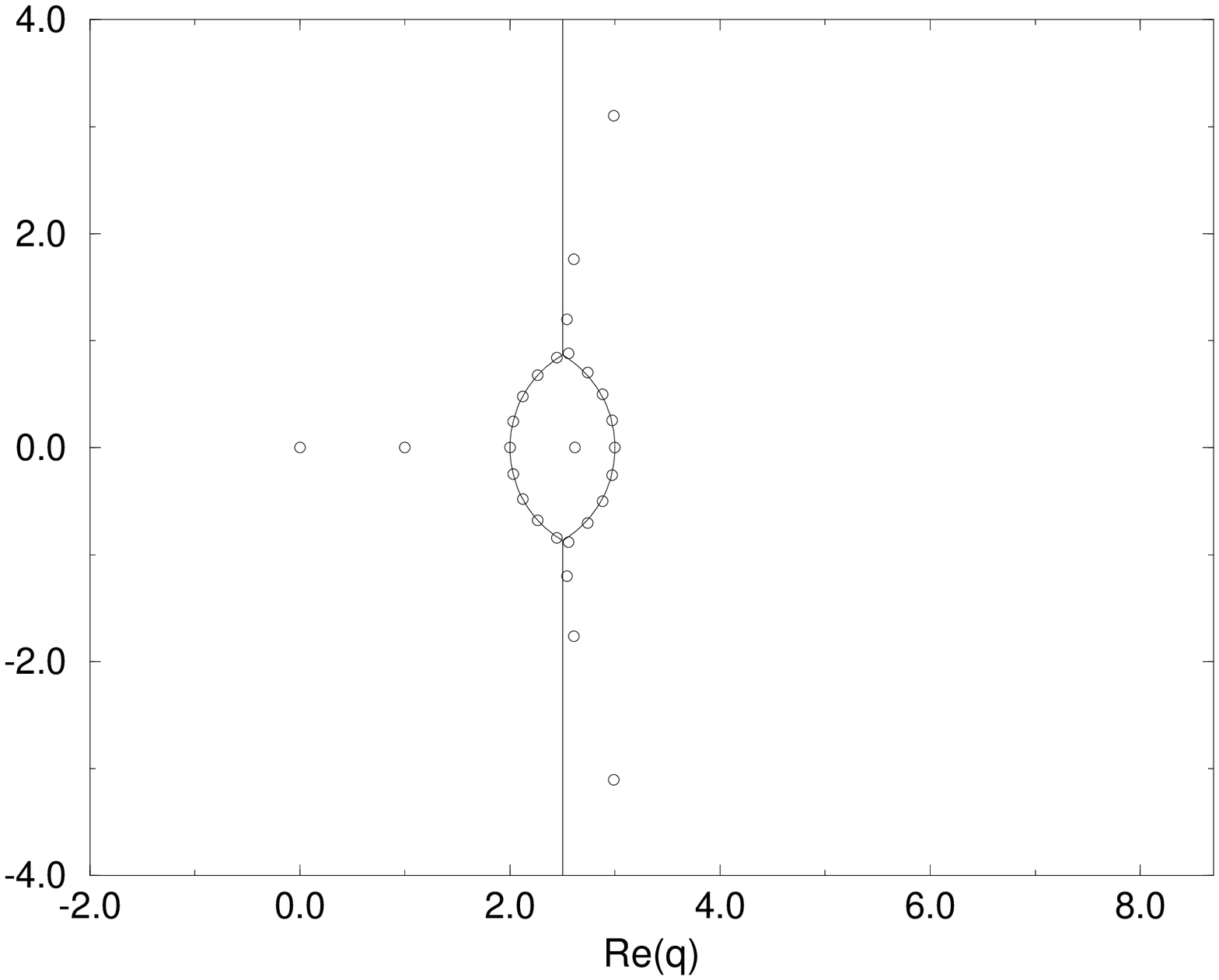}
\caption{Diagram showing regional boundaries comprising ${\cal B}$ for 
$W(\{B\},q)$ for bipyramid graphs and zeros of $P(B_n,q)$ for $n=29$.} 
\label{bipyramidzerosfig}
\end{figure}

   The bipyramid $B_n$ is formed from the biwheel $U_n$ by removing the bond 
connecting the adjoined vertex to the center vertex of the biwheel.  A
bipyramid graph can be inscribed on the 2-sphere $S^2$ and, in this sense, can
be considered to be 2-dimensional. The chromatic polynomial for $B_n$ is 
\beq
P(B_n,q) = q \Bigl \{ (q-2)^{n-2} + (q-1)(q-3)^{n-2} + (-1)^n(q^2-3q+1) 
\Bigr \}
\label{pbn}
\eeq
Here we find a more complicated diagram describing the analytic structure (see
also Ref. 13); this is shown in Fig. \ref{bipyramidzerosfig} and consists of 
three regions: 
\beq
R_1: Re(q) > \frac{5}{2} \quad {\rm and} \quad |q-2| > 1
\label{br1}
\eeq
\beq
R_2: Re(q) < \frac{5}{2} \quad {\rm and} \quad |q-3| > 1
\label{br2}
\eeq
and 
\beq
R_3: |q-2| < 1 \quad {\rm and} \quad |q-3| < 1
\label{br3}
\eeq
The boundaries between these regions are thus the two circular arcs 
\beq
{\cal B}(R_1,R_3): 
q = 2+e^{i\theta} \ , -\frac{\pi}{3} < \theta < \frac{\pi}{3}
\label{leftarc}
\eeq
and 
\beq
{\cal B}(R_2,R_3): q=3+e^{i\phi} \ , \frac{2\pi}{3} < \phi < \frac{4\pi}{3}
\label{rightarc}
\eeq
together with the semi-infinite vertical line segments 
\beq
{\cal B}(R_1,R_2) = \{q\}: \quad 
Re(q)=\frac{5}{2} \quad {\rm and} \quad |Im(q)| > \frac{\sqrt{3}}{2}
\label{lines}
\eeq
These meet at the intersection points $q=5/2 \pm i\sqrt{3}/2$.  We find that 
\beq
W(\{B\},q) = q-2 \quad {\rm for} \ \ q \in R_1 
\label{wbr1}
\eeq
For the other regions, we have, in general, 
\beqs
|W(\{B\},q)| = & = & |q-3| \quad {\rm for} \ \ q \in R_2 \cr
               & = & 1     \quad {\rm for} \ \ q \in R_3 
\label{wb}
\eeqs
With specific choices of $1/n$'th roots, one can choose $W(\{B\},q)=q-3$ in
$R_2$ and $-1$ in $R_3$. 

  This example provides an illustration of the noncommutativity of limits 
(\ref{wnoncomm}) for a real non-integer point $q_0$, namely, 
$q_0 = (1/2)(3 + \sqrt{5}) = Be_5 = 2.618..$, where the $r$'th Beraha number 
$Be_r$ is given by \cite{beraha}
\beq
Be_r = 4 \cos^2 (\pi/r)
\label{be}
\eeq
for $r=1,2,..$. The point $Be_5$ 
lies in the region $R_3$ where $q^2-3q+1$ is the dominant term 
and is one of the two roots of this polynomial (the other root lies in region
$R_2$ and hence plays no role in $W(\{B\},q)$).  
In Fig. \ref{bipyramidzerosfig} we have also plotted zeros of the bipyramid
chromatic polynomial $P(B_n,q)$ for a typical finite $n=29$.  These will be
discussed in Section 5.

   Since $W(\{G\},q)$ is bounded above by $q$, it is common to remove this
factor and define a reduced function 
\beq
W_r(\{G\},q) = q^{-1}W(\{G\},q)
\label{wr}
\eeq
which has a finite limit as $|q| \to \infty$.  There have been a number of
calculations of Taylor series expansions in the variable $1/(q-1)$ for 
functions equivalent to $W_r$ in the case where $G$ is a regular lattice; see,
for example, Ref. \cite{kewser} (and earlier references therein) 
for the square, triangular, and honeycomb lattices.  Clearly, these series
expansions rely on the property that, for these lattices, 
$W_r(\{G\},q)$ is an analytic function in the $1/q$ plane at the origin,
$1/q=0$.  This analyticity of $W_r(\Lambda,q)$ at $1/q=0$ is proved by exact
results for the triangular lattice and is strongly supported by numerical
calculations of zeros of $P(\Lambda,q)$ for $\Lambda=sq,hc$ \cite{baxter87},
which show that the respective regional boundaries for these three lattices are
compact, and do not extend to infinite distance from the origin in the complex
$q$ plane.  However, our exact result for the infinite-$n$ bipyramid function 
$W(\{B\},q)$ and its region diagram demonstrates that, in general, the 
infinite-$n$ reduced limit $W_r(\{G\},q)$ of chromatic polynomials for a 
given family of graphs 
$\{G\}$ is {\it not} guaranteed to be analytic at $1/q=0$: in the case of the
bipyramid graphs, the portion of the regional boundary ${\cal B}$ comprised 
by the line segment (\ref{lines}) runs vertically through the origin of the 
$1/q$ plane, and $W_r(\{B\},q)$ is not analytic at $1/q=0$.

\subsection{Cyclic Ladder Graphs $L_{2n}$ }

   The cyclic ladder graphs with $2n$ vertices can be visualized as two
$n$-circuit graphs (rings) $C_n$, one above the other, with the $i$'th vertex
of one $n$-circuit connected by a vertical bond to the $i$'th vertex of the 
other $n$-circuit.  The chromatic polynomial for this family of graphs was 
calculated in Ref. \cite{bds} (where they are called prism graphs): 
\beq
P(L_{2n},q) = (q^2-3q+3)^n +(q-1) \Bigl \{ (3-q)^n + (1-q)^n \Bigr \} 
+ q^2-3q+1
\label{pl2n}
\eeq

\begin{figure}
\epsfxsize=3.5in
\epsffile{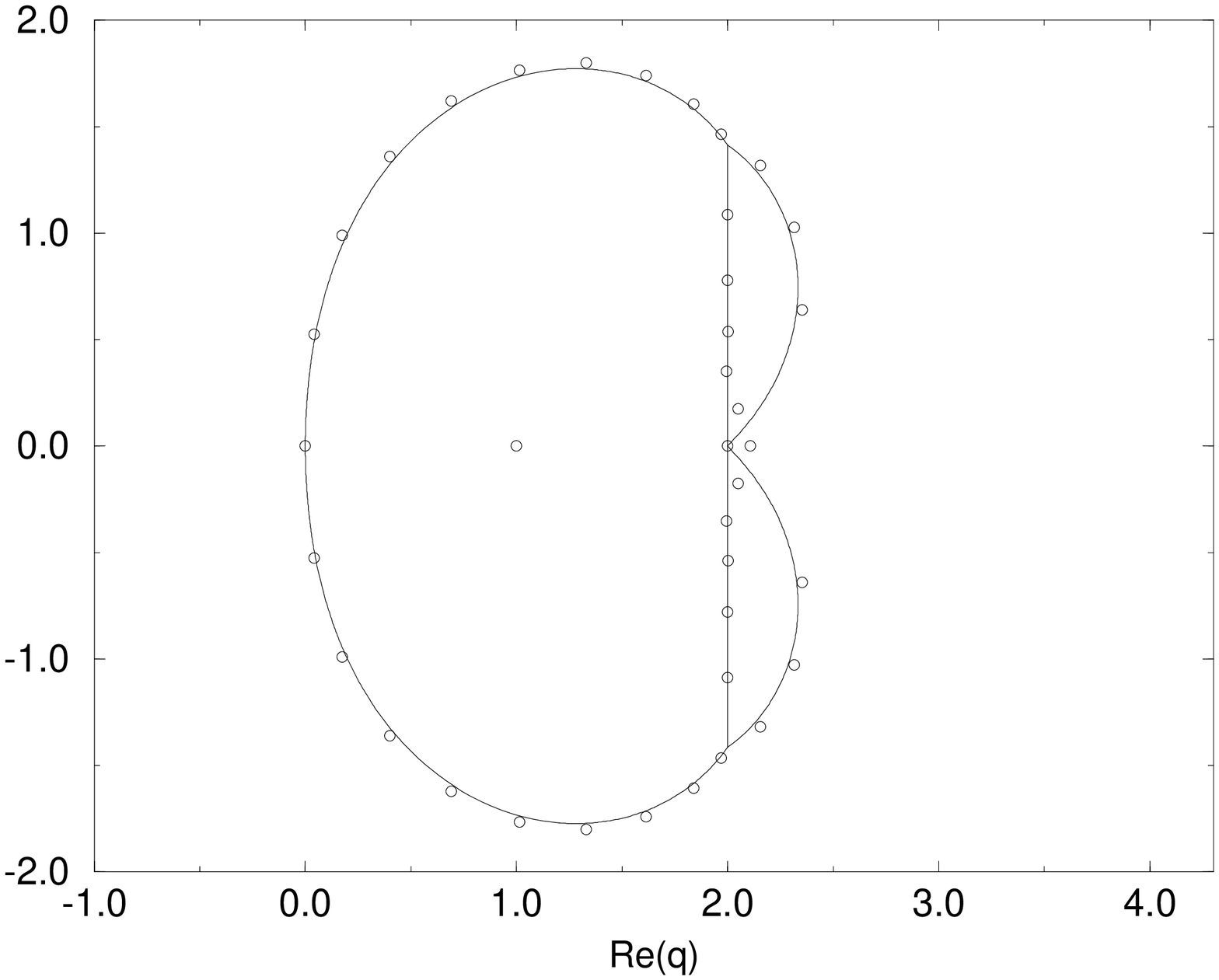}
\caption{Diagram showing regional boundaries comprising ${\cal B}$ for
$W(\{L\},q)$ for cyclic ladder graphs and zeros of $P(L_{2n},q)$ for $2n=38$.}
\label{ladderzerosfig}
\end{figure}

 From this we compute the region diagram shown in Fig. \ref{ladderzerosfig}, 
consisting of four 
regions: $R_1$, $R_2$, $R_2^*$, and $R_3$ in which, respectively, 
(1) $(q^2-3q+3)^n$, (2)-(2)$^*$ $(1-q)^n$, and (3) $(3-q)^n$ are the leading 
terms.  We find 
\beq
W(\{L\},q) = q^2 -3q + 3 \quad {\rm for} \quad q \in R_1
\label{wlr1}
\eeq
\beq
|W(\{L\},q)| = |1-q| \quad {\rm for} \quad q \in R_2 \ \ {\rm or} \ \ R_2^* 
\label{wlr2}
\eeq
\beq
|W(\{L\},q)| = |3-q| \quad {\rm for} \quad q \in R_3
\label{wlr3}
\eeq
Formally, one can choose $1/n$'th roots so that $W(\{L\},q)=1-q$ and $3-q$ in
the respective regions ($R_2$,$R_2^*$) and $R_3$. 
Note that, in contrast to the situation with the bipyramid graphs $B_n$, there
is no region where all $|a_j(q)| < 1$ so that the $c_0(q)$ term (which is equal
to $q^2-3q+1$ here) never dominates. 
The boundary between regions $R_2$, $R_2^*$, and $R_3$ is the line 
$|1-q|=|3-q|$ where these are both leading terms, which is comprised of 
the line segments 
\beq
{\cal B}(R_2,R_3) = \{q\}: \quad Re(q) = 2 \ , \quad 0 < Im(q) < \sqrt{2}
\label{bndyr2r3}
\eeq
and its complex conjugate for ${\cal B}(R_2^*,R_3)$. 
Similarly, the boundary separating $R_1$ from $R_3$ is the locus of solutions
to the degeneracy condition $|q^2-3q+3|=|3-q|$ where these are leading terms. 
This boundary runs vertically through the origin $q=0$ and extends over, on the
right, to two complex-conjugate triple points $q = 2 \pm \sqrt{2}i$
where it meets the vertical line boundary (\ref{bndyr2r3}) and its complex
conjugate.  Similarly, the boundaries separating $R_1$ from $R_2$ and $R_1$
 from $R_2^*$ are comprised of the locus of solutions to the degeneracy 
condition $|q^2-3q+3|=|1-q|$ where these are leading terms; as shown in
Fig. \ref{ladderzerosfig}, these boundaries 
extend from the above triple points over to a four-fold intersection point at
$q=2$. In passing, we note that our region diagram differs from that reported 
in Ref. \cite{bds}, where the right-most curves were thought to terminate and
hence not completely separate from $R_1$ the two additional regions that we
have identified as $R_2$ and $R_2^*$. 
In Fig. \ref{ladderzerosfig}, we also show zeros of $P(L_{2n},q)$ for $2n=38$. 
We shall discuss these in Section 5. 

\subsection{Twisted Ladder (M\"obius) Graphs} 

   One may also consider ladder graphs the ends of which are twisted once 
before being joined; these graphs are denoted twisted ladder or M\"obius
graphs, $M_{2n}$.  (It is easy to see that if one twists the ends an even
number of times, this is equivalent to no twist, and any odd 
number of twists are equivalent to a single twist.)  The chromatic polynomial
is the same as that for $L_{2n}$ except for the $c_0(q)$ term \cite{bds}
\beq
P(M_{2n},q) = (q^2-3q+3)^n +(q-1) \Bigl \{ (3-q)^n - (1-q)^n \Bigr \} - 1
\label{pm2n}
\eeq
Since there is no region where the constant term $c_0(q)$ (equal to $-1$ here)
is dominant, we find that 
\beq
W(\{M\},q) = W(\{L\},q)
\label{wm}
\eeq

\section{Theorem for Zeros of Chromatic Polynomials for Certain $\{G\}$ }

   A general question that one may ask about zeros of chromatic polynomials is
whether all, or some subset, of the zeros for an $n$-vertex graph $G$ in the 
family $\{G\}$ lie exactly on the boundary curves ${\cal B}$.  One knows that
as $n \to \infty$, aside from the discrete general set of zeros of $P(G,q)$,
viz., $q_0 = 0,1$, and, for graphs containing one or more triangles, $q=2$, the
remainder of the zeros merge to form the union of boundaries ${\cal B}$
separating various regions in the complex $q$ plane.  (Some of the set 
$\{q_0\}$ may also lie on ${\cal B}$.)  However, the fact that the zeros move
toward, and merge to form, this boundary ${\cal B}$ in the $n \to \infty$ limit
does not imply that, for finite graphs $G$, some subset of zeros will lie 
precisely on ${\cal B}$.  We have investigated this question and have found
that there do exist some families of graphs $\{G\}$ for which the zeros of 
$P(G,q)$ (aside from certain members of the set $\{q_0\}$) lie exactly 
on the respective boundary curves ${\cal B}$.  We shall present a theorem 
and proof on this.  Interestingly, we find that in all such cases, 
${\cal B}$ consists of a unit 
circle centered at a certain integral point on the positive real $q$ axis.  
We emphasize, however, that this type of behavior is special and is not shared
by other families of graphs that we have studied.  Furthermore, for the
families of graphs for which the theorem does hold, the positions of the unit 
circles differ for different $\{G\}$.  Finally, as we shall show, the zeros
populate the full circle with constant density.

   We find that the theorem applies for the following three families of graphs:
(i) cyclic; (ii) wheel, and (iii) biwheel.  We begin with the cyclic graphs.
The form of $P(C_n,q)$ differs depending on whether $n$ is even, say $n = 2m$
or odd, say $n = 2m+1$.  For even $n \ge 4$, we calculate the factorization
\beq
P(C_{n=2m},q) = q(q-1)\prod_{j=0}^{m-2}
\Bigl \{ q-(1+e^\frac{(2j+1)\pi i}{n-1}) \Bigr \}
\Bigl \{ q-(1+e^\frac{-(2j+1)\pi i}{n-1}) \Bigr \}
\label{pcnevenfactors}
\eeq
and for odd $n \ge 5$, 
\beq
P(C_{n=2m+1},q) = q(q-1)(q-2)\prod_{j=1}^{m-1}
\Bigl \{ q-(1+e^\frac{2j\pi i}{n-1}) \Bigr \}
\Bigl \{ q-(1+e^\frac{-2j\pi i}{n-1}) \Bigr \}
\label{pcnoddfactors}
\eeq
Special cases for lower $n$ are $P(C_2,q)=q(q-1)$ and 
$P(C_3,q)=q(q-1)(q-2)$.

   For the wheel graphs, for odd $n \ge 5$, we find the factorization
\beq
P((Wh)_{n=2m+1},q) = q(q-1)(q-2)\prod_{j=0}^{m-2}
\Bigl \{ q-(2+e^\frac{(2j+1)\pi i}{n-2}) \Bigr \}
\Bigl \{ q-(2+e^\frac{-(2j+1)\pi i}{n-2}) \Bigr \}
\label{pwhnoddfactors}
\eeq
and for even $n \ge 6$, 
\beq
P((Wh)_{n=2m},q) = q(q-1)(q-2)(q-3)\prod_{j=1}^{m-2}
\Bigl \{ q-(2+e^\frac{2j\pi i}{n-2}) \Bigr \}
\Bigl \{ q-(2+e^\frac{-2j\pi i}{n-2}) \Bigr \}
\label{pwhnevenfactors}
\eeq
Special cases for lower $n$ are $P((Wh)_3,q)=q(q-1)(q-2)$ 
and $P((Wh)_4,q)=q(q-1)(q-2)(q-3)$.

For the biwheel graphs we calculate for even $n \ge 6$, 
\beq
P(U_{n=2m},q) = q(q-1)(q-2)(q-3)\prod_{j=0}^{m-3}
\Bigl \{ q-(3+e^\frac{(2j+1)\pi i}{n-3}) \Bigr \}
\Bigl \{ q-(3+e^\frac{-(2j+1)\pi i}{n-3}) \Bigr \}
\label{punevenfactors}
\eeq
and for odd $n \ge 7$, 
\beq
P(U_{n=2m+1},q) = q(q-1)(q-2)(q-3)(q-4)\prod_{j=1}^{m-2}
\Bigl \{ q-(3+e^\frac{2j\pi i}{n-3}) \Bigr \}
\Bigl \{ q-(3+e^\frac{-2j\pi i}{n-3}) \Bigr \}
\label{punoddfactors}
\eeq
Special cases for lower $n$ are $P(U_4,q)=q(q-1)(q-2)(q-3)$ and 
$P(U_5,q)=q(q-1)(q-2)(q-3)(q-4)$. 

 These factorizations constitute a proof of the following

\begin{flushleft}
Theorem 2
\end{flushleft}
Except for isolated zeros at $q=1$ for
$C_n$, at $q=0,2$ for $(Wh)_n$, and at $q=0,1,3$ for $U_n$, the zeros of 
$P(C_n,q)$, $P((Wh)_n,q)$, and $P(U_n,q)$  all lie on the respective unit 
circles $|q-q_\odot|=1$ where $q_\odot(C_n)=1$, $q_\odot(Wh_n)=2$, and 
$q_\odot(U_n)=3$. 
Furthermore, the zeros are equally spaced around the respective
unit circles, and in the $n \to \infty$ limit, the density $g(\{G\},\theta)$ of
zeros on the respective circles $q = q_\odot + e^{i\theta}$, $-\pi < \theta \le
\pi$, is a constant, independent of $\theta$.  If one normalizes $g$ 
according to
\beq
\int_{-\pi}^{\pi} g(\{G\},\theta) \ d\theta = 1
\label{gnorm}
\eeq
then
\beq
g(\{G\},\theta) = \frac{1}{2\pi} \quad {\rm for} \quad \{G\} = \{C\}, \ \ 
\{Wh\}, \ \ \{U\} 
\label{gg}
\eeq

  For the cyclic ladder and twisted ladder graphs $L_{2n}$ and $M_{2n}$, we 
find the type of behavior which occurs with complex-temperature zeros of spin 
models: the zeros lie close to, but not, in general, precisely on, the 
asymptotic boundaries ${\cal B}$.  This is illustrated by the plots of zeros 
of $P(L_{38},q)$ in Fig. \ref{ladderzerosfig}.  As one also finds in
calculations of complex-temperature zeros in statistical mechanical spin 
models (see, e.g., \cite{abe,ih}), the densities of zeros along certain 
boundary curves are very small; in Fig. \ref{ladderzerosfig} 
this occurs on ${\cal B}(R_1,R_2)$ near 
the intersection point $q=2$.  Similar low densities of zeros were observed for
$P(tri,q)$ on the boundary near $q=0$ and the right-most boundary near $q=4$
\cite{baxter87}.  In statistical mechanics, the density of zeros $g$ near a
critical point $z_c$ behaves as 
\beq
g \sim |z-z_c|^{1-\alpha'}
\label{gz}
\eeq
where $\alpha'$ denotes the critical exponent describing the (leading) 
singularity in the specific heat at $z=z_c$ as one approaches this point from
within the broken-symmetry phase: $C_{sing} \sim |z-z_c|^{-\alpha'}$
\cite{abe}. Equivalently, the (leading) singularity in the free
energy at $z=z_c$ is given by $f_{sing} \sim |z-z_c|^{2-\alpha'}$. (Similar
statements apply for the approach to $z_c$ from within the symmetric phase with
the replacement $\alpha' \to \alpha$.)  Analogously, in the present context,
the density of zeros of $P(G,q)$ for a finite-$n$ graph $G$ near a singular 
point is determined by the nature of the singularity in the asymptotic 
function $W(\{G\},q)$: if one denotes the singularity in the 
function $\ln W(\{G\},q)$ at a point $q_c$ as 
\beq
\ln W(\{G\},q)_{sing} \sim |q-q_c|^{2-\alpha_\phi'}
\label{ssing}
\eeq
where $\alpha_\phi'$ will, in general, depend on the direction ($\phi$) of 
approach to $q_c$, then the corresponding density of zeros of $P(\{G\},q)$ 
as one approaches this point is 
\beq
g(\{G\},q) \sim |q-q_c|^{1-\alpha_\phi'}
\label{gp}
\eeq

For the bipyramid, as is evident in Fig. \ref{bipyramidzerosfig}, one sees 
that, except for the general zeros at $q=0$ and 1 and a zero very near to
$q=Be_5 = 2.618...$, the inner zeros 
do lie near to the arcs forming the boundaries ${\cal B}(R_1,R_3)$ and 
${\cal B}(R_2,R_3)$, but the outer zeros do not lie very close to the line 
segments of ${\cal B}(R_1,R_2)$, given by eq. (\ref{lines}), and only approach
these line segments slowly as $n$ increases.  
Since this latter behavior only occurs for the part of ${\cal B}$ extending 
to $q=5/2 \pm i \infty$, it is plausible that it may be connected with the 
fact that this component of the boundary is noncompact.  
This inference is also consistent with the fact that of the families of 
graphs which we have studied, the bipyramid graphs form the only
family with a noncompact ${\cal B}$ and the only family for which we have
observed this deviation. 

\section{Theorem on Singular Boundary of $W(\Lambda,q)$}

    In recent work \cite{ih}-\cite{yy} on complex-temperature and Yang-Lee 
(complex-field) singularities of Ising models, it has been quite fruitful to 
carry out a full complexification of both the temperature-dependent Boltzmann 
weight $u = z^2 = e^{-4K}$ and the field-dependent Boltzmann weight
$\mu=e^{-2\beta H}$ and
to study the singularities in the two-dimensional $C^2$ manifold depending 
on $(z,\mu)$ or $(u,\mu)$.  This approach unifies the previously separate
analyses of complex-temperature and Yang-Lee singularities; one sees that a
given singular point $(z_c,\mu_c)$ or $(u_c,\mu_c)$ in the $C^2$ manifold 
manifests itself as a singular point in the complex $z$ or $u$ plane for a
fixed $\mu$ and equivalently as a singular point in the complex $\mu$ plane for
fixed $z$ or $u$.  

    We find this approach to be equally powerful here. Starting from the
relation (\ref{zprel}) between the chromatic polynomial and the $T=0$ Potts
antiferromagnet on a graph $G$, we consider the two-dimensional complex
manifold $C^2$ spanned by $(a,q)$ (where $a$ was defined in eq. (\ref{a})). For
sufficiently large $q$, namely, $q > 2\zeta$, where $\zeta$ is the coordination
number of the lattice $G = \Lambda$, the Dobrushin theorem implies \cite{sokal}
that the Potts antiferromagnet is disordered, with exponential decay of
correlation functions, at $T=0$.  As one decreases $q$, the AFM ordering
tendency of the system increases, and, as $q$ decreases through a critical
value depending on the dimensionality $d$ and lattice $\Lambda$, the model can
become critical at $T=0$, or equivalently, $K = -\infty$.  As one decreases $q$
further, the AFM critical temperature increases from zero to positive values
(i.e. $K_c$ increases from $-\infty$ to a finite negative value).  The critical
value $q_c$ thus 
separates two regions in $q$: (i) the $q > q_c$ region, where the
system is disordered at $T=0$ and (ii) an interval of $q < q_c$ where the
system has AFM long-range order at $T=0$ (and for a finite interval $0 \le T
\le T_c$, where $K_c = J/(k_BT_c)$).  Now, using the relations (\ref{zprel}),
(\ref{fwrel}) and making the projection from the $(a,q)$ space onto the real
$a$ axis, just as the disordered, $Z_N$-symmetric phase of the Potts AF
must be separated by a non-analytic phase boundary from the broken-symmetry 
phase with AFM long-range order \cite{pb}, so also, making the projection
onto the real $q$ axis for the $W(\Lambda,q)$ function, it follows that the 
range $q > q_c$ and an adjacent interval $q < q_c$ must be separated by a 
non-analytic boundary.  Furthermore, just as, by analytic continuation, the 
complex-temperature extension of the disordered phase of the Potts
antiferromagnet must be completely separated by a non-analytic phase boundary
 from the complex-temperature extension of the antiferromagnetically ordered
phase, so also the region in the complex $q$ plane containing the line segment
$q > q_c$ must be completely separated from the region containing the adjacent
interval to the left of $q_c$ in this plane.  
Since the zero-temperature criticality of the Potts AF
and the critical value $q_c$ are both projections of the
singular point $(a_c=0, q_c)$ in the $C^2$ manifold, we have derived the
following theorem:

\vspace{2mm}

\begin{flushleft}

Theorem 3: 
For a given lattice $\Lambda$, the point $q_c$ at which the right-most
region boundary for $W(\Lambda,q)$ crosses the real $q$ axis corresponds to the
value of $q$ at which the critical point $a_c$ of the Potts 
antiferromagnet on this lattice first passes through zero as one decreases $q$
 from large positive values. 

\vspace{4mm} 

This point $q_c$ is the maximal finite real point of non-analyticity of
$W(\Lambda,q)$. 

\end{flushleft}

We now discuss the application of this theorem to three specific 2D lattices. 
For this purpose, we recall that the $q$-state Potts model can be defined for
non-integral as well as integral values of $q$ because of the equivalent
representation of the partition function \cite{kf},\cite{wurev}, 
\cite{baxter73} 
\beq
Z = \sum_{G' \subseteq G}v^{b(G')}q^{n(G')}
\label{zpequiv}
\eeq
where $G'$ denotes a subgraph of $G = \Lambda$, $v=(a-1)$, $b(G')$ is the
number of bonds and $n(G')$ the number of connected components of $G'$. (Recall
that one can see the connection of this with (\ref{zprel}) by taking the 
$K \to -\infty$ ($v \to -1$) limit of (\ref{zpequiv}), which yields the 
Whitney expression for $P(G,q)$ \cite{whit}.) 

\subsection{$q_c$ for the Honeycomb Lattice}

   For the honeycomb (hc) lattice, the 
paramagnetic-ferromagnetic (PM-FM) and PM-AFM critical points are both
determined by the equation \cite{kj}
\beq
q^2 + 3q(a-1)-(a-1)^3=0
\label{hccriteq}
\eeq
As $q$ decreases in the range from 4 to $q=Be_5=2.618$, one of the roots of 
eq. (\ref{hccriteq}) increases from $-1$ to 0.  This root can be identified as
the AFM critical point $a_c(q)$ by going in the opposite direction, 
increasing $q$ from its Ising value, $q=2$ and tracking $a_c(q)$, which
decreases from $a_c(2)=2-\sqrt{3}$ to $a(q=Be_5)=0$, where the AFM phase is
squeezed out and there is no longer any finite-temperature AFM critical point,
which now occurs only at $T=0$.  Hence, our theorem implies that 
\beq
q_c(hc) = \frac{3+\sqrt{5}}{2}= Be_5 = 2.618..
\label{qchc}
\eeq
i.e., this is the value of $q$ where the right-most regional boundary of 
$W(hc,q)$ crosses the real axis in the complex $q$ plane \cite{p3afhc}.  
We may compare this with the numerical
calculation of zeros of $P(hc,q)$, as a function of $q$, on finite honeycomb 
lattices in Ref. \cite{baxter87}.  For this comparison, we first note that 
 from the analogous study of the triangular lattice \cite{baxter87}, 
one sees the crossing point of a boundary curve increases by 
about $\Delta q \simeq 0.4$ from an $8 \times 8$ triangular lattice with 
cylindrical boundary conditions (CBC's) to the thermodynamic limit. Assuming
that a similar finite-size shift occurs for the honeycomb lattice, and noting
that the curve of zeros calculated on the $8 \times 8$ hc lattice with CBC's
cross the real axis at $q \simeq 2.2$, we find that these numerical results are
consistent with the result of Theorem 3 for the thermodynamic limit. 

\subsection{$q_c$ for the Square Lattice}

   For the square lattice, the PM-AFM critical point of the Potts
antiferromagnet is given by \cite{baxter82}
\beq
(a+1)^2 = 4-q
\label{afmcurve}
\eeq
i.e., 
\beq
a_c(sq) = -1 + \sqrt{4-q}
\label{acq}
\eeq
As $q$ decreases from 4 to 3, this value of $a_c$ increases from $-1$ to 0.  
Hence, 
\beq
q_c(sq)=3
\label{qcsq}
\eeq
Our Theorem 3 
then identifies $q_c(sq)$ as the point where the right-most region
boundary of $W(sq,q)$ crosses the real axis in the complex $q$ plane.  Using
the same rough estimate for the finite-size shift between the $8 \times 8$ 
square lattice with CBC's and the thermodynamic limit as was observed for the
triangular lattice, viz., $\Delta q \simeq 0.4$ and 
noting that a curve of zeros calculated for this finite square lattice 
crosses the real axis at $q \simeq 2.6$ \cite{baxter87}, we see that our
inference (\ref{qcsq}) is consistent with the numerical calculations in
Ref. \cite{baxter87}. 

\subsection{$q_c$ for the Triangular Lattice}

   Baxter's exact solution for $W(tri,q)$ \cite{baxter87} shows that in this 
case,
\beq
q_c(tri) = 4
\label{qctri}
\eeq
where the right-most boundary, ${\cal B}(R_1,R_2)$, crosses 
the real $q$ axis.  Theorem 3 implies that the two other singular points 
where the boundaries ${\cal B}(R_2,R_3)$ and ${\cal B}(R_3,R_1)$ cross this
axis, at $q=3.82...$ and $q=0$, respectively, also correspond to singular
points of the Potts antiferromagnet in the $a$ plane at these two $q$ values. 

\section{Numerical Calculations of $W(\Lambda,q)$ for $\Lambda=sq,hc,tri$}

  The effect of ground state disorder and associated nonzero ground state
entropy $S_0$ has been a subject of longstanding interest.  A physical
example is ice, for which $S_0 = 0.82 \pm 0.05$ cal/(K-mole), i.e.,
$S_0/k_B = 0.41 \pm 0.03$ \cite{ice,liebwu}. In statistical mechanical models,
such a ground state entropy may occur in contexts such as the Ising (or
equivalently, $q=2$ Potts) antiferromagnet on the triangular \cite{wannier} 
or kagom\'e \cite{knsuto} lattices, where there is frustration.  However,
ground state entropy can also occur in what is arguably a simpler context: 
one in which
it is not accompanied by any frustration.  On a given lattice, for sufficiently
large $q$, Potts antiferromagnets exhibit ground state entropy without 
frustration; restricting to integral values of $q$, this is true for 
$q \ge 3$ on the square and honeycomb lattices, and for $q \ge 4$ on the
triangular lattice \cite{frust}.   For the given range
of $q$ on the respective lattices, since the internal energy $U$ approaches 
its $T=0$ value $U(T=0) = -J\langle \delta_{\sigma_i \sigma_j} \rangle_{T=0} =
0$ exponentially fast, it follows that for $T \to 0$, i.e., $K \to -\infty$,
$\lim_{K \to -\infty} \beta U = 0$.  Hence, from the general relation 
$S = \beta U + f$, it follows that, for this range of values of $q$, the ground
state entropy (per site) and reduced free energy for the Potts antiferromagnet
are related according to 
\beq
S_0(\Lambda,q) = f(\Lambda,q,K=-\infty) = \ln W(\Lambda,q) 
\label{sp}
\eeq
and hence $\ln W(\Lambda,q)$ is a measure of this ground state entropy. 
Accordingly, it is of interest to calculate $W(\Lambda,q)$ for various lattices
$\Lambda$ and values of $q$.  Moreover, from a mathematical point of view, 
for positive integer
$q$, the numerical calculation of $W(\Lambda,q)$ gives an accurate measure of
the asymptotic growth of $P(\Lambda,q) \sim W(\Lambda,q)^n$ as the number of
lattice sites $n \to \infty$.  

\begin{figure}
\epsfxsize=3.5in
\epsffile{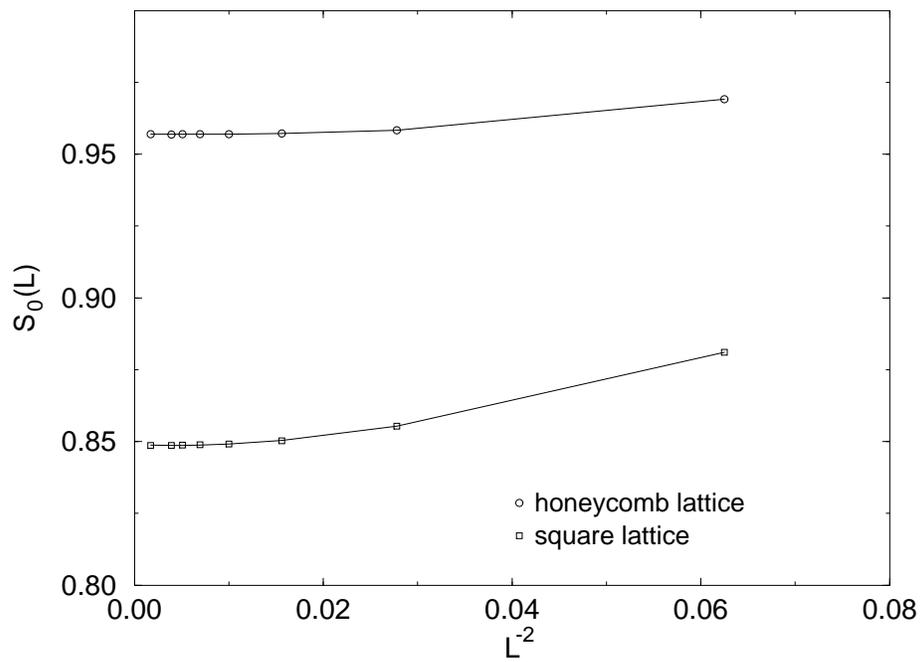}
\caption{
Measurements of ground state entropy $S_0$, as a function of lattice
size, for the $q=4$ Potts AF on the honeycomb and square lattices.} 
\label{s0fitfig}
\end{figure}

   Extending our earlier calculation of $W(hc,3)$ 
\cite{p3afhc}, we have calculated $W(hc,q)$ for integer $4 \le q \le 10$.  We 
use the relation for the entropy 
\beq
S(\beta) = S(\beta=0) + \beta U(\beta) - \int_0^{\beta} U(\beta')d\beta'
\label{seq}
\eeq
which is known to provide a very accurate method for calculating $S_0$
\cite{binder}. We start the integration at $\beta=0$ with $S(\beta=0)=\ln q$ 
for the $q$-state Potts antiferromagnet and utilize a Metropolis algorithm 
with periodic BC's for several $L \times L$ lattices with the length $L$
varying over the values 4,6,8,10,12,14,and 16 for all cases, and up to $L=24$
for certain cases.  Since $U(K)$ very rapidly approaches 
its asymptotic value of 0 as $K$ decreases
past about $K=-5$, the RHS of (\ref{seq}) rapidly approaches a
constant in this region, enabling one to obtain the resultant value of
$S(\beta=\infty)$ for each lattice size.  For each value of $q$, we then 
perform a least squares 
fit to this data and extrapolate the result to the thermodynamic 
limit, and then obtain $W$ from (\ref{sp}).  We use double precision arithmetic
for all of our computations. Typically, we ran several thousands sweeps through
the lattice for
thermalization before calculating averages. Each average was calculated
using between 9,000 and 20,000 sweeps through the lattice.  As we have
discussed in Ref. \cite{p3afhc}, for $ q \ge 3$ on the honeycomb lattice 
(and also for $q \ge 4$ on square lattice), the finite-size dependence of 
$S_0$ is not simply of the form
$S_0(\Lambda; L \times L, q) = S_0(\Lambda,q) + c_{\Lambda,1}^{(q)}L^{-2}$; we
fit our measurements with an empirical function of the form 
\beq
S_0(\Lambda; L \times L, q) = S_0(\Lambda,q) + c_{\Lambda,1}^{(q)}L^{-2} + 
c_{\Lambda,2}^{(q)}L^{-4} + c_{\Lambda,3}^{(q)}L^{-6}
\label{s0fit}
\eeq
As an example, we show in Fig. \ref{s0fitfig} the ground state entropy as a 
function of 
$L^{-2}$, for the case $q=4$ on the square and honeycomb lattices.  As a check,
we have confirmed that our measurements yield numbers consistent with the exact
result $S_0=0$ for $q=2$ and $\Lambda=hc,sq$. 

\begin{table}
\begin{center}
\begin{tabular}{|c|c|c|c|c|} \hline \hline & & & & \\
$q$ & $W(hc,q)$ & $hc$ series & $W(sq,q)$ & $sq$ series \\
& & & & \\
\hline \hline
3  &  1.6600(5)     &  1.6600  &  1.53965(45) &  1.5396007..  \\
4  &  2.6038(7)     &  2.6034  &  2.3370(7)   &  2.3361  \\
5  &  3.5796(10)    &  3.5795  &  3.2510(10)  &  3.2504  \\
6  &  4.5654(15)    &  4.5651  &  4.2003(12)  &  4.2001  \\
7  &  5.5556(17)    &  5.5553  &  5.1669(15)  &  5.1667  \\
8  &  6.5479(20)    &  6.5481  &  6.1431(20)  &  6.1429  \\
9  &  7.5424(22)    &  7.5426  &  7.1254(22)  &  7.1250  \\
10 &  8.5386(25)    &  8.5382  &  8.1122(25)  &  8.1111  \\
\hline
\end{tabular}
\end{center}
\caption{Values of $W(\Lambda,q)$ for $\Lambda = hc,sq$ and $3 \le q \le 10$ 
 from Monte Carlo measurements, compared with large-$q$ series.  The entry for
$W(sq,3)$ is from the exact expression. See text for further details.} 
\label{wtable}
\end{table}

\begin{figure}
\epsfxsize=3.5in
\epsffile{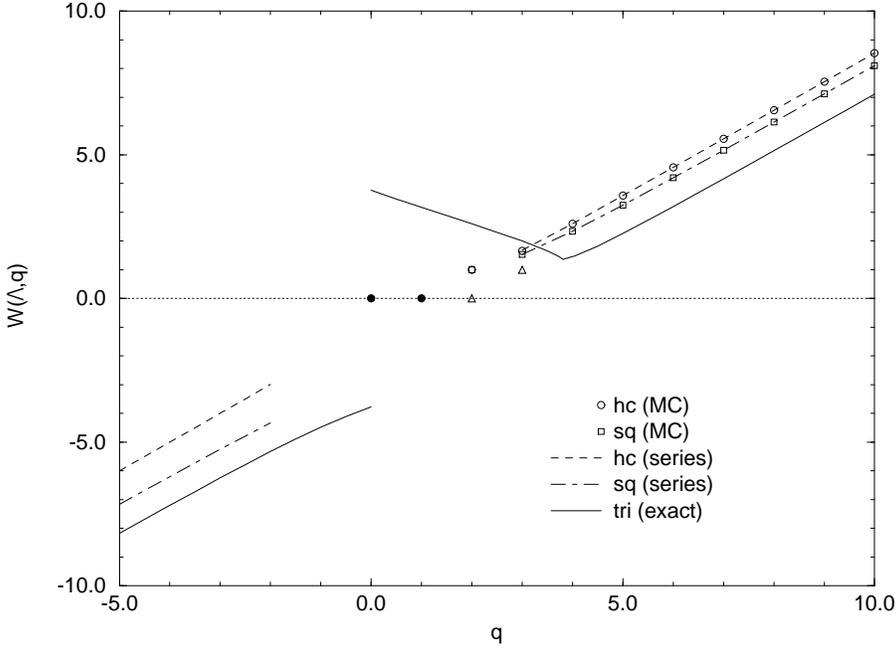}
\caption{Plot of $W(\Lambda,q)$ for $\Lambda=sq,hc,tri$.  At the special points
$q_s=0,1$ for all lattices, the zero values (denoted by the symbol $\bullet$) 
apply for the order of 
limits in the definition $D_{nq}$ in eq. (\ref{wdefnq}).  For the triangular 
lattice at the special points $q_s=2,3$, the respective values 0 and 1
(denoted by the symbol $\triangle$) also apply for this ordering of limits 
(\ref{wdefnq}), as does 
the value $W(\Lambda,2)_{D_{nq}}=1$ for $\Lambda=sq,hc$. The $\Lambda=tri$ 
curve is plotted for the definition $D_{qn}$ in eq. (\ref{wdefqn}).}
\label{wallfig}
\end{figure}

In the course of these calculations, we have obtained a measurement of
$W(hc,3)$ which is more accurate than, and in excellent agreement with, the 
value that we reported recently in Ref. \cite{p3afhc} (where we quoted the
uncertainty very conservatively).  This improvement is due 
to our use of larger lattices, up to $24 \times 24$, and double precision
arithmetic, in the present work.  Our results for $W(hc,q)$ are presented 
in Table \ref{wtable} (with conservatively estimated uncertainties given in 
parentheses) and plotted in Fig. \ref{wallfig}, together with the respective 
values obtained by evaluating the large-$q$ series. 
For a lattice $\Lambda$, this series has the form 
\beq
W(\Lambda,q) = q\Bigl ( \frac{q-1}{q} \Bigr )^{\zeta/2}\overline W(\Lambda,q)
\label{wseriesdef}
\eeq
where, as above, $\zeta$ is the lattice coordination number, and 
\beq
\overline W(\Lambda,q)=1+\sum_{n=1}^\infty w_n y^n \ , \quad y = \frac{1}{q-1}
\label{wseries}
\eeq
For the honeycomb lattice, $\overline W(hc,q)^2 = 1 + y^5 + 2y^{11} + 4y^{12} +
...$, calculated through $O(y^{18})$ \cite{kewser}.  Because of the sign 
changes in the hc series (the coefficients of the first five terms are 
positive, while those of the remaining four terms are negative),
it is difficult to make a reliable extrapolation.  
Accordingly, for Table \ref{wtable} and Fig. \ref{wallfig} we simply 
use a direct evaluation of the sum.  As is evident from this figure, the
agreement with our Monte Carlo calculation is excellent.  From our result
(\ref{qchc}) above, it follows that the large-$q$ series cannot be applied 
below $q=2.62$, and we have plotted it only down to the
integer value, $q=3$.  For both the honeycomb and square lattices, we
also show the exact results $W(\Lambda,2)_{D_{nq}}=1$ for $\Lambda=hc,sq$ (as a
superimposed circle and square) and
$W(\Lambda,0)_{D_{nq}}=W(\Lambda,1)_{D_{nq}}=0$ (as a dot $\bullet$), but we 
emphasize that these values assume the order of the limits in the definition 
$D_{nq}$ in eq. (\ref{wdefnq}) and the respective values calculated with the 
other order of 
limits in the definition $D_{qn}$ in eq. (\ref{wdefqn}) for these lattices
(which values are not known exactly) could well be different from these
$D_{nq}$ values. 

   It is also of interest to compare $W(\Lambda,q)$ for the other two regular
2D lattices, square and triangular.  Unlike the honeycomb lattice, there have
been previous Monte Carlo measurements for the square lattice
\cite{chenpan} for lattice sizes between $3 \times 3$ and $7 \times 7$ and $q$
values up to 10.  We have extended these to considerably larger lattices, 
including $16 \times 16$ for all $q$ values. 
As a check, for $q=3$, from calculations on $L \times L$ lattices 
with periodic boundary conditions, for $L=4,6,8,10,12,14$ and 16, we obtain 
the fit 
\beq
S_0(sq,3)=0.431556 + 1.095289 L^{-2} 
\label{wsq3form}
\eeq
This yields an asymptotic value which is in excellent agreement with the exact
result \cite{lieb} $S_0(sq,3) = (3/2)\ln(4/3) = 0.43152311...$ (i.e., 
$W(sq,3) = 1.5396007...$);
\beq
\frac{|S_0(sq,3)_{exact}-S_0(sq,3)_{MC}|}{S_0(sq,3)_{exact}} = 
0.76 \times 10^{-4}
\label{scomp}
\eeq
The coefficient of the $L^{-2}$ term agrees with a previous determination for
this $q=3$ case \cite{wsk}.  Our fitting procedure for $q \ge 4$ has been
discussed in conjunction with eq. (\ref{s0fit}) above. We also compare our 
Monte Carlo calculations with the large-$q$ series, (\ref{wseriesdef}),
(\ref{wseries}) which, for $\Lambda=sq$ was calculated to $O(y^{18})$ in
Ref. \cite{kewser}.  The agreement is again excellent. In passing, we note that
a calculation of the series for $\overline W(sq,q)$ to order $O(y^{36})$ 
has been reported in Ref. \cite{bakaev}, but we have checked that additional 
terms in this longer series have a negligible effect in the
comparison of the series with our numerical results.  Given our result that 
$q_c(sq)=3$ in (\ref{qcsq}), the large-$q$ series cannot be applied for 
$q < 3$.  It is interesting that the agreement between the series and our
measurements is quite good even down to the respective region boundaries which
we have deduced at $q_c(hc)=2.618$ and $q_c(sq)=3$. This suggests that the
non-analyticities at these respective points on the honeycomb and square
lattices are evidently not so strong as to cause the series to deviate strongly
 from the actual values of $W(\Lambda,q)$. 

In passing, we remark that of course our results are consistent with the
following rigorous bounds: (i) the general upper bound $W(\Lambda, q) < q$;
(ii) the upper bound for the square lattice \cite{biggsbound}, 
$W(sq,q) \le (1/2)(q-2+\sqrt{q^2-4q+8})$,
which is more restrictive than (i) for $q > 1/2$; (iii) the lower bound
applicable for any bipartite lattice, $W(\Lambda_{bip.},q) \ge \sqrt{q-1}$; and
(iv) the lower bound for the square lattice \cite{biggsbound},
$W(sq,q) \ge (q^2-3q+3)/(q-1)$.  Note that for $q=2$, both lower bounds (iii)
and (iv) are realized as equalities, $W(sq,2)=1$.  Although there is a range of
$q$ above 2 where (iv) lies below (iii), for $q \ge 3$, (iv) lies above (iii),
i.e. is more restrictive.  Some recent rigorous upper bounds on $P(G,q)$ for
general $G$ have been given in Ref. \cite{pbounds}, but these only improve 
the prefactor $A$ multiplying $P(G,q) \le Aq^n$, and hence still 
yield $W(G,q) \le q$, as in (i).

   In the case of the triangular lattice, since, to our knowledge, there is no
numerical evaluation in the literature of the expressions for $W(tri,q)$ given
in Ref. \cite{baxter87}, we have carried this out and plotted the resultant
function in Fig. \ref{wallfig}.  The point $q=3$ is an example of a
special point $q_s$ discussed in section 2, where the behavior of $P(G,q)$
changes abruptly from (\ref{pgasym}) to (\ref{pgqs}) and where, consequently,
the two limits in (\ref{wnoncomm}) do not commute.  As noted above, since there
are just 6 ways of coloring a triangular lattice (equivalently, the ground
state of the Potts AF is 6-fold degenerate), $W(tri,3)_{D_{nq}}=1$.  We have
indicated this with a symbol $\triangle$ in Fig. \ref{wallfig}. However,
with the other order of limits, (\ref{wdefqn}), $\lim_{q \to 3}W(tri,q) =
W(tri,3)_{D_{qn}} \ne 1$.  (The actual value is $W(tri,3)_{D_{qn}}=2.$) 
In Fig. \ref{wallfig}, at the other special points $q_s=0,1$ (indicated with 
$\bullet$) and $q_s=2$ (indicated with $\triangle$) we have shown the values 
$W(tri,q)_{D_{nq}}=0$.  Again, however, because of the noncommutativity of
limits in
(\ref{wnoncomm}) as discussed in section 2, the value of $W(tri,q)_{D_{qn}}$
calculated with the other order of limits, (\ref{wdefqn}) is nonzero.  At
$q=1,2$, it is positive, while at $q=0$, the function has a discontinuity
involving a flip in sign:
\beq
\lim_{q \to 0^-} W(tri,q) = - \lim_{q \to 0^+} W(tri,q)
\label{wdiscq0}
\eeq
The sign of $W(tri,q)$ for negative real $q$ is unambiguous, since 
this interval is part of the region $R_1$ and where consequently, there is a
clear choice of $1/n$'th root, given by $r=0$, in (\ref{pphase}).  This is 
also clear from the large-$q$ series. 
In Refs. \cite{baxter87}, it was noted that the transfer matrix 
calculation used there can fail at the Beraha numbers $q = Be_r$; from the
discussion that we have given in section 2 of this paper, we would view this as
a specific realization of the general noncommutativity of limits 
(\ref{wnoncomm}). 

   A general property one observes in Fig. \ref{wallfig} is that for these
three lattices, for a fixed value of $q$ in the 
range $q \ge 4$, $W(\Lambda,q)$ is a monotonically decreasing function of the
lattice coordination number $\zeta$ and a monotonically increasing
function of the lattice ``girth'' $\gamma$, defined \cite{graphs} as the 
number of bonds, or equivalently, vertices contained in a minimum-distance 
circuit. Here, $\zeta=3,4,6$ and $\gamma=6,4,3$ for $\Lambda=hc,sq,tri$. The
dependence on the girth is easily understood: the smaller the girth, the more 
stringent is the constraint that no two colors on adjacent vertices can be 
the same.  Concerning the dependence on $\zeta$, we note that for tree graphs,
where one can vary $\zeta$ for fixed $\gamma$ ($\gamma=\infty$), $P(T_n,q)$ and
hence $W(\{T\},q)$ are actually independent of $\zeta$ (c.f. eq. (\ref{wt})). 
This is also true for the magnitude $|W(\Lambda,q)|$
for real negative $q$. 

   For the bipartite (square and honeycomb) lattices, we 
observe the following general trend: as $q$ increases from $q=3$, 
the ratios 
\beq
r_W(\Lambda,q) = \frac{W(\Lambda,q)}{W(\Lambda,q)_{max}} = 
 \frac{W(\Lambda,q)}{q}
\label{rw}
\eeq
and the related 
\beq
r_S(\Lambda,q) = \frac{S(\Lambda,q;T=0)}{S(\Lambda,q,T=\infty)} = 
\frac{\ln \Bigl ( W(\Lambda,q) \Bigr )}{\ln q}
\label{rs}
\eeq
are monotonically increasing functions of $q$.  The ratio $r_S(\Lambda,q)$ 
has a physical
interpretation as measuring the residual disorder present in the $q$-state
Potts antiferromagnet at $T=0$, relative to its value at $T=\infty$.  This 
ratio is substantial; for example, from Table \ref{wtable}, one sees that 
$r_S(hc,3)=0.461$ and $r_S(sq,3)=(3/2)\ln(4/3)/\ln 3 = 0.393$ while 
$r_S(hc,10)=0.931$ and
$r_S(sq,10)=0.909$.  On the triangular lattice, as $q$ increases from $q=4$,
one again finds that $r_W$ and $r_S$ monotonically increase; for example, 
$r_S(tri,4) = 0.273$ while $r_S(tri,10) = 0.852$.

Returning to the square and honeycomb lattices, although we cannot use our 
Monte Carlo method to evaluate $W(hc,q)$ for
negative $q$, we can use the large-$q$ series, since the negative $q$ axis is
in the region $R_1$.  Of course these
series cannot be used all the way in to $q=0$; in Fig. \ref{wallfig}, we plot
them up to $q=-2$.

\section{Conclusions}
    
   In conclusion, we have presented some results on the analytic properties
of the asymptotic limiting function $W(\{G\},q)$ obtained from the chromatic 
polynomial $P(G,q)$.  We have pointed out that the formal equation (\ref{w})
is not, in general, sufficient to define the function $W(\{G\},q)$ because of
the noncommutativity of limits (\ref{wnoncomm}) at certain special points, 
and we
have provided the necessary clarification for a complete definition of this
asymptotic function.  Using mathematical results on chromatic polynomials for
several families of graphs $\{G\}$, we have calculated $W(\{G\},q)$ exactly 
for these families.  From these results, we have determined the non-analytic 
boundaries separating various regions in the complex $q$ plane for each of the
$W(\{G\},q)$.  We have also studied the zeros of chromatic polynomials for 
these families of graphs and have proved a theorem stating that for some 
families, all but a
finite set of these zeros lie exactly on certain unit circles centered at
positive integer points on the real $q$ axis.  Using the connection of
chromatic polynomials to the partition function of the $q$-state Potts
antiferromagnet on a lattice $\Lambda$ at $T=0$, in 
conjunction with a generalization to both complex
$q$ and complex temperature, we have presented another theorem specifying the
position of the maximal (finite) real point $q_c(\Lambda)$ where 
$W(\{G\}=\Lambda,q)$ is non-analytic and have applied this
to determine $q_c$ on the square and honeycomb lattices.  Finally, we have
given Monte Carlo measurements of $W(hc,q)$ (and $W(sq,q)$) for integral 
$3 \le q \le 10$ and compared these with large-$q$ series. 
Our results illustrate the fascinating
and deep connections between the mathematics of chromatic polynomials and 
their limits on the one hand, and the statistical mechanics of 
antiferromagnetic Potts models on the other.

This research was supported in part by the NSF grant PHY-93-09888.

\vfill
\eject

\begin{thebibliography}{99}
\bibitem{bc}{As is well-known, boundary conditions do not affect the 
thermodynamic limit except in the sense that they can affect which of 
degenerate ground states the system picks in the broken-symmetry phase.}
\end{thebibliography}
\end{document}